 \documentclass[final,3p,times]{elsarticle}

\usepackage{amssymb}
\usepackage{amsmath}

\usepackage{lineno}

\usepackage{float}
\usepackage{subcaption}
\usepackage{wasysym}
\usepackage{ulem}

\usepackage{url}
\usepackage[breaklinks]{hyperref}

\usepackage{xcolor}

\journal{Nuclear Instruments and Methods in Physics Research A}

\begin{document}

\begin{frontmatter}



\title{Application of a high-precision distributed uranium source for determining the effective mass and volume of a HPGe detector} 

\author[2]{A. Barabash}
\author[1]{S. Evseev}
\author[1]{D. Filosofov}
\author[1,3,6]{V. Kazalov}
\author[1,4]{T. Khussainov\corref{cor1}}
\author[1,3,5]{A. Lubashevskiy}
\author[1]{N.D. Mokhine}
\author[1,3,5]{D. Ponomarev}
\author[1]{S. Rozov}
\author[1]{S. Vasilyev}
\author[1]{M. Vorobyeva}
\author[1]{E. Yakushev}
\author[2]{V. Yumatov}

\cortext[cor1]{Corresponding author. Email: khusainov@jinr.ru}


\affiliation[2]{organization={National Research Center «Kurchatov Institute», Kurchatov Complex of Theoretical and Experimental Physics}, 
            city={Moscow},
            postcode={117218},
            country={Russia}}

\affiliation[1]{organization={Dzhelepov Laboratory of Nuclear Problems, Joint Institute for Nuclear Research},
            addressline={6 Joliot-Curie}, 
            city={Dubna},
            postcode={141980},
            country={Russia}}

\affiliation[3]{organization={Institute for Nuclear Research of the Russian Academy of Sciences},
            addressline={7a Prospect 60-letiya Oktyabrya}, 
            city={Moscow},
            postcode={117312},
            country={Russia}}

\affiliation[6]{organization={Kabardino-Balkarian State University named after H.M. Berbekov},
addressline={173 Ulitsa Chernyshevskogo}, 
city={Nalchik},
postcode={360004},
country={Russia}}

\affiliation[4]{organization={Institute of Nuclear Physics of the Ministry of Energy of the Republic of Kazakhstan},
            addressline={1 Ibragimov Street}, 
            city={Almaty},
            postcode={050032},
            country={Kazakhstan}}
            
\affiliation[5]{organization={Lebedev Physical Institute of the Russian Academy of Sciences},
            addressline={53 Leninskiy Prospect}, 
            city={Moscow},
            postcode={119991},
            country={Russia}}

\begin{abstract}
A distributed uranium source with the accurately certified activity of $^{238}$U has been used to verify the effective mass of the HPGe detector intended for the $\nu$GeN neutrino experiment. The source, dissolved in nitric acid, provides homogeneous irradiation of the detector crystal allowing the study of its mass and volume. The experimental spectra obtained with the distributed source have been compared to the detailed Geant4 Monte Carlo simulations. The measured counting rates of several $\gamma$-lines agree with the simulated efficiencies, confirming that the detector's mass and volume coincide with the manufacturer’s specifications. The results demonstrate the applicability of such sources for mass calibration of HPGe detectors.

\end{abstract}


\begin{keyword}

HPGe detector \sep uranium calibration source \sep Geant4 simulation \sep $\nu$GeN experiment



\end{keyword}

\end{frontmatter}



\section{Introduction}
The study of neutrino properties is an important area in modern elementary particle physics. Coherent elastic neutrino--nucleus scattering (CE$\nu$NS) is the most probable process of neutrino interaction with matter in the low-energy neutrino region. The $\nu$GeN experiment \cite{alekseev2022first} at the Kalinin Nuclear Power Plant is aimed at searching for coherent elastic scattering of reactor neutrinos off germanium nuclei. The detection of CE$\nu$NS requires a large neutrino flux, a low background, large target mass, and a low energy threshold. For an unbiased interpretation of the experimental results, all these parameters must be known with a high precision.

In this paper, the target mass, i.e. the effective mass of germanium, is under study. Routinely, in order to check the germanium detector mass indicated by the manufacturer, the detectors are scanned either with point-like sources, like in the GERDA experiment \cite{agostini2015production} where the $^{241}$Am and $^{60}$Co sources were used, or with X-ray radiography \cite{maidana2013efficiency}. Recently, mainly due to the demands of ICP-MS \cite{qiao2018application}, a variety of samples distributed in solutions have become commercially available. Among them, liquid samples with dissolved uranium are especially advantageous for different calibration purposes in nuclear spectroscopy. The macro quantities of $^{238}$U have relatively low radioactivity, $12440 \pm 12$ Bq/g \cite{NNDC}. Due to this fact, such volume sources can have a precisely determined activity, highly suitable for calibration of the absolute detector efficiency. In addition, their distributed nature allows simultaneous irradiation of the crystal with $\gamma$-rays from multiple directions, enabling the study of both the detector mass and the volume. This makes such a source advantageous compared to point-like sources whose activity is typically determined with a precision of 3--7\% \cite{Ritverc}. However, the deviations from the exact point-like geometry and possible non-uniformity of the activity may introduce additional uncertainties.

In this paper, we used the uranium source for the verification of the effective mass of the HPGe detector provided by the manufacturer. The measurements performed in the presence of the uranium sample have been compared to a computational model.

\section{Source description}
The uranium calibration source distributed in nitric acid solution was produced by Inorganic Ventures \cite{IVcert}. The source parameters are given in Table \ref{tab:source}.

\begin{table}[h]
\centering
\caption{Parameters of uranium calibration source specified by manufacturer \cite{IVcert}.}
\label{tab:source}
\begin{tabular}{ll}
\hline
Parameter & Value \\
\hline
Volume (cylindrical bottle) & 125 mL \\
Density & 1.010 g/mL \\
Volume concentration HNO$_3$ & 2\% \\
Uranium concentration & 1000 $\pm$ 5 $\mu$g/mL \\
\hline
\end{tabular}
\end{table}

\begin{figure}[H]
\centering
\includegraphics[width=1\linewidth]{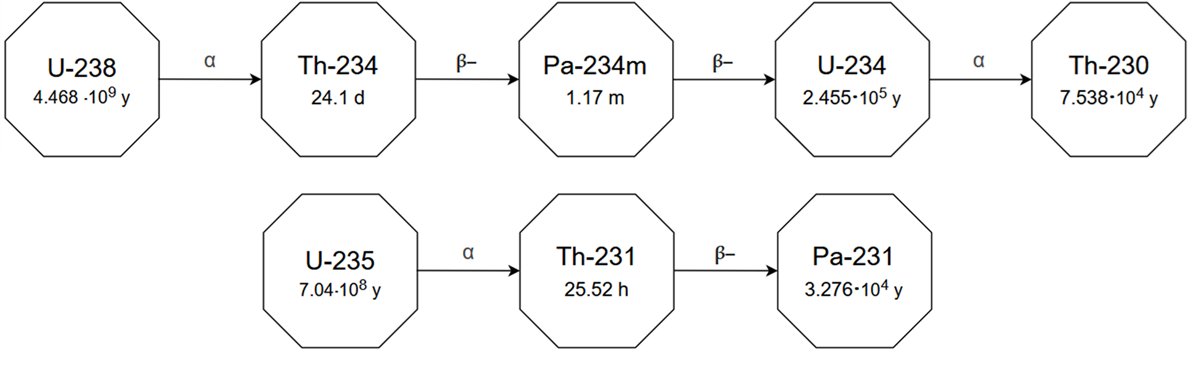}
\caption{Simplified decay schemes of $^{238}$U and $^{235}$U \cite{NNDC}.}
\label{fig:decay}
\end{figure}

The specified mass of dissolved uranium is 125 mg, i.e. the $^{238}$U activity in the sample is 1555 $\pm$ 8 Bq. Thus, the activity is known with a precision of 0.5\%. According to the manufacturer, the uranium sample is depleted in $^{235}$U. The certified abundances of $^{238}$U and $^{235}$U are $99.8 \pm 0.1$\% and $0.19 \pm 0.05$\%, respectively \cite{IVcert}. The solution is guaranteed to be homogeneous with a minimum sample size of 0.2 mL. The source was produced in August 2022, and the measurements were carried out in January 2024, ensuring that the $^{238}$U decay chain (Figure \ref{fig:decay}) was in equilibrium up to $^{234m}$Pa. Similarly, in the $^{235}$U decay chain, equilibrium was established up to $^{231}$Th.

\section{Experimental setup and Monte Carlo simulations}
The experimental setup was constructed at DLNP of JINR at sea level. The detector under study is a point-contact low-background high-purity germanium detector manufactured by Mirion Technologies. According to manufacturer specification the HPGe crystal is a cylinder of 70 mm in diameter and 70 mm in height, dead layer of the crystal is less than 0.5 mm and active volume is 268 cm$^3$. The manufacturer does not provide the accuracy with which the crystal’s dimensions are determined, so it can be assumed that they may vary slightly for each specific crystal. The detector belongs to the same production batch as the one used in the $\nu$GeN experiment \cite{alekseev2022first}.

To suppress the external background, the detector is surrounded by passive and active shielding. The passive shielding consists of three layers: 10 cm of copper, 8 cm of borated polyethylene, and 10 cm of lead. The active shielding is a muon veto made of the 5-cm-thick plastic scintillator which fully encloses the passive shielding. Before the shielding installation, the detector’s energy scale was calibrated using ambient $\gamma$-lines.

\begin{figure}[H]
\centering
\includegraphics[width=0.65\linewidth]{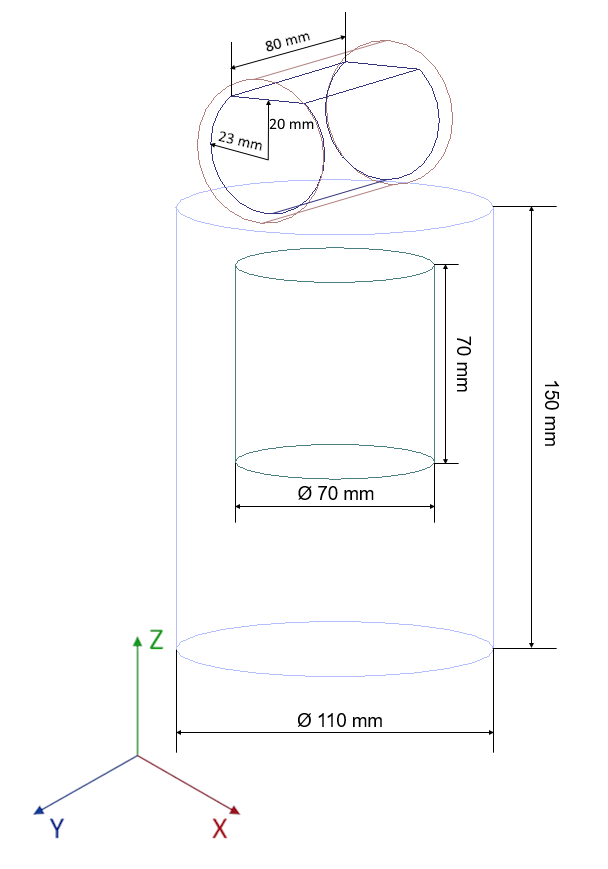}
\caption{Visualization of uranium source placement on detector's endcap.}
\label{fig:setup}
\end{figure}

The geometry of the detector within the shielding and the uranium source on the detector's endcap (Figure \ref{fig:setup}) was reproduced in Geant4.10.04.p03 \cite{collaboration2003geant4}. Since the exact internal geometry of the detector is unknown two boundary conditions of the crystal dimensions and the dead layer thickness were set according to the manufacturer's specifications. Both had an active volume of 268 cm$^3$ (which corresponds to an effective mass of 1.42 kg): first was \diameter 70 mm $\times$ 70 mm with 0.06 mm dead layer and second was \diameter 70.9 mm $\times$ 70.9 mm with 0.5 mm dead layer. In the simulation, $1.36\times 10^{9}$ decays of $^{238}$U and its daughters down to $^{234m}$Pa were generated by using the Geant4 Shielding Physics List.

 Shifting the source in the MC model by 10--15 mm changed the counting rate by less than 2\%. Since the real positioning is known with a precision of a few mm, that systematic error can be considered negligible.

\section{Measurements}
\begin{figure}[h]
\centering
\includegraphics[width=0.75\linewidth]{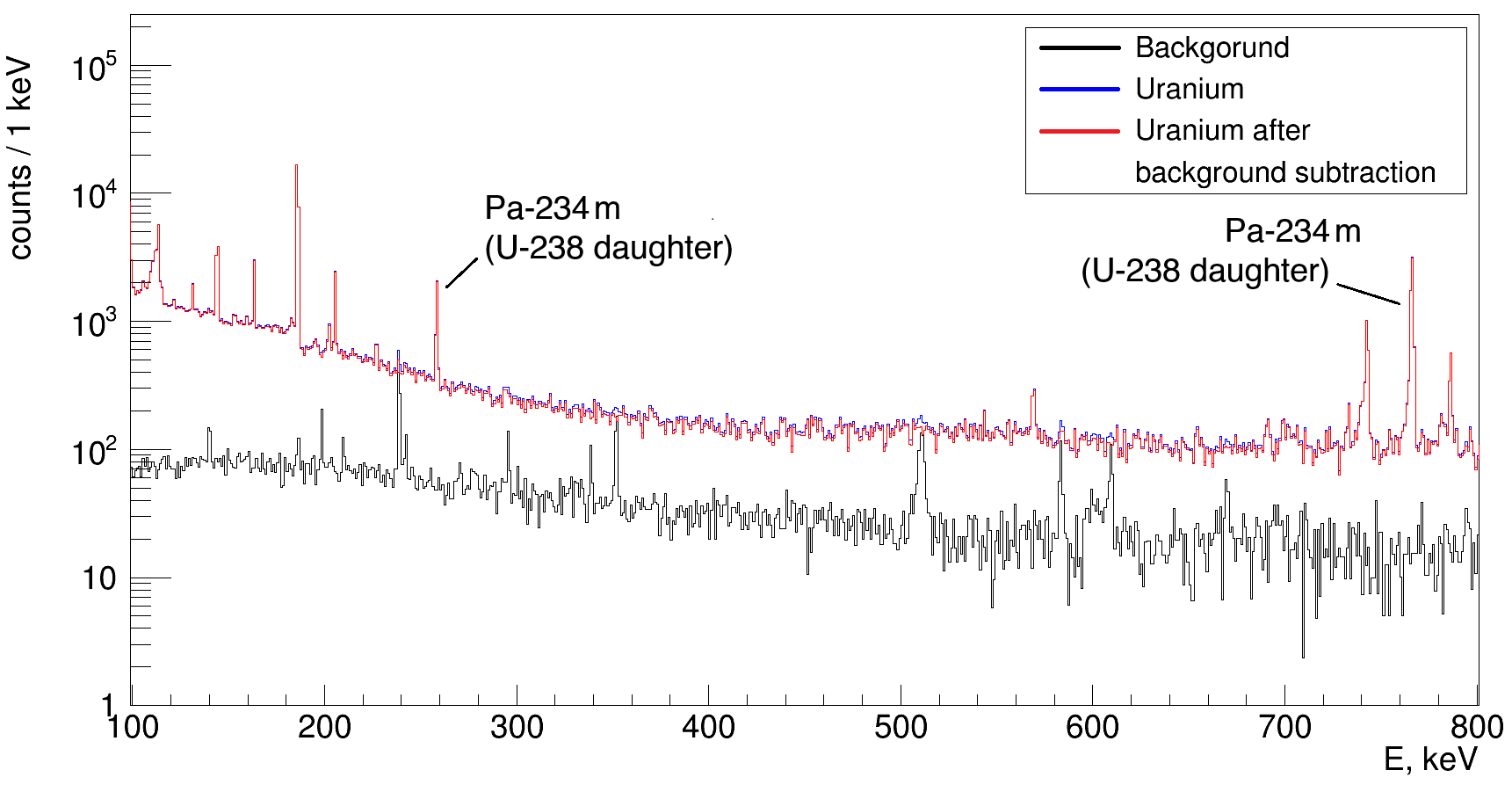}
\caption{Experimental spectra normalized to 24.4 h. Black: background measurements; blue: uranium source measurements; red: uranium after background subtraction.}
\label{fig:spectra}
\end{figure}

The background conditions were measured over 10 days, and the measurements with the uranium source lasted 24.4 hours. The experimental spectra obtained are depicted in Figure \ref{fig:spectra}. The background contribution to the uranium spectrum is only about 1\% at low energies and up to 10\% at higher energies.

The detector is optimized for measurements in the low-energy region. Its upper operational energy range is limited to 800 keV.  Within this range, two $\gamma$-ray lines with energies of 766 and 258 keV were selected. 

\section{Results}
\begin{figure}[H]
\centering
\includegraphics[width=0.75\linewidth]{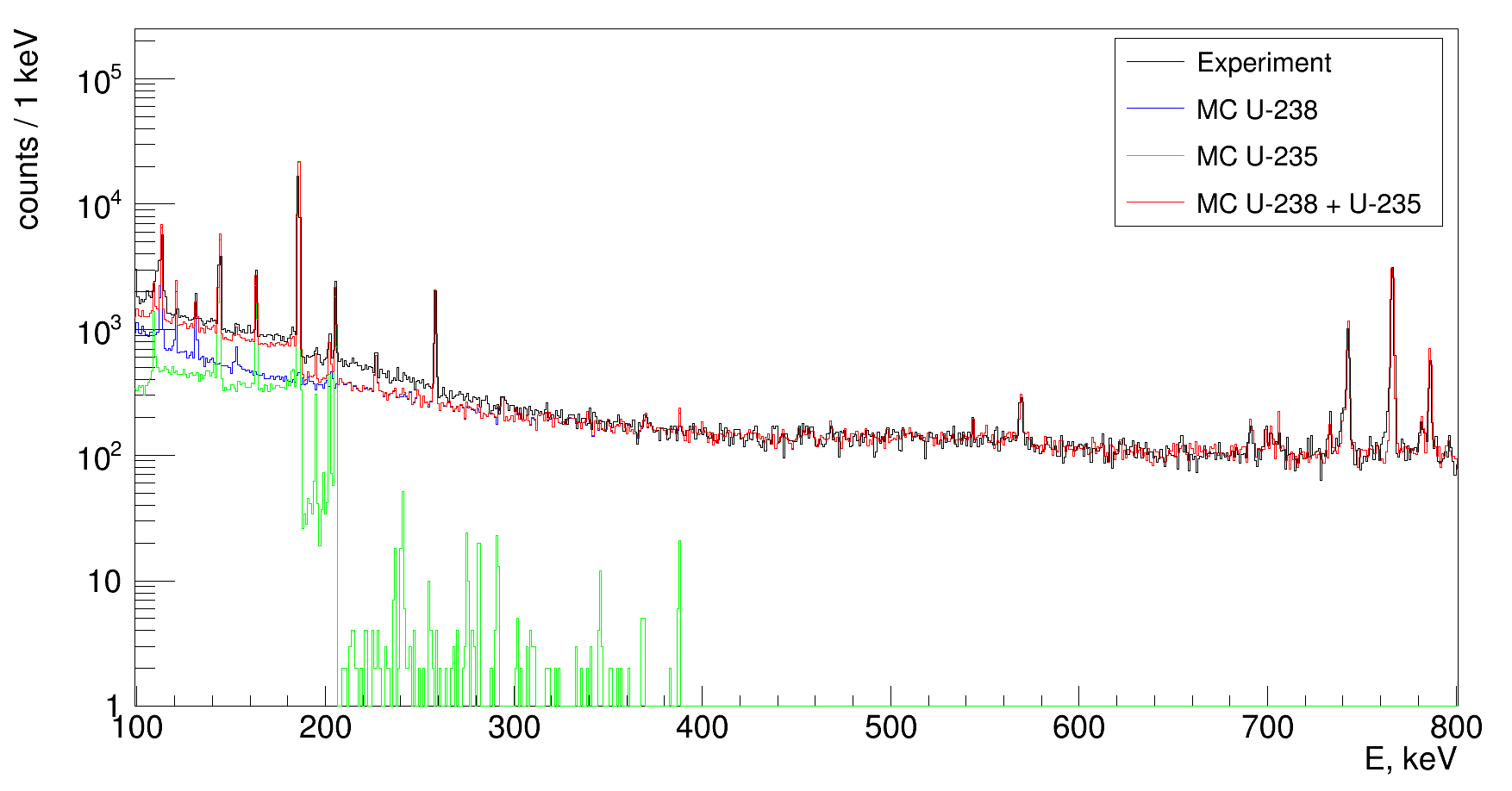}
\caption{Comparison of experimental spectrum (black) and Geant4 simulations: $^{238}$U (blue), $^{235}$U (green), $^{238}$U+$^{235}$U (red).}
\label{fig:comparison}
\end{figure}

Figure \ref{fig:comparison} shows the experimental spectrum and separate yields from the Geant4 spectra of $^{238}$U, $^{235}$U, and the total MC spectrum. There are some discrepancies between the experimental data and the Monte Carlo simulations in the region below 300 keV. Possible origins of such discrepancies are: non-precise knowledge of internal geometry of the detector, large relative error in the $^{235}$U isotope content and uncertainties in the modeling of physical processes in Geant4, such as bremsstrahlung.

\begin{figure}[H] 
  \centering
  \begin{subfigure}{0.45\textwidth}
  \centering
    \includegraphics[width=0.93\linewidth]{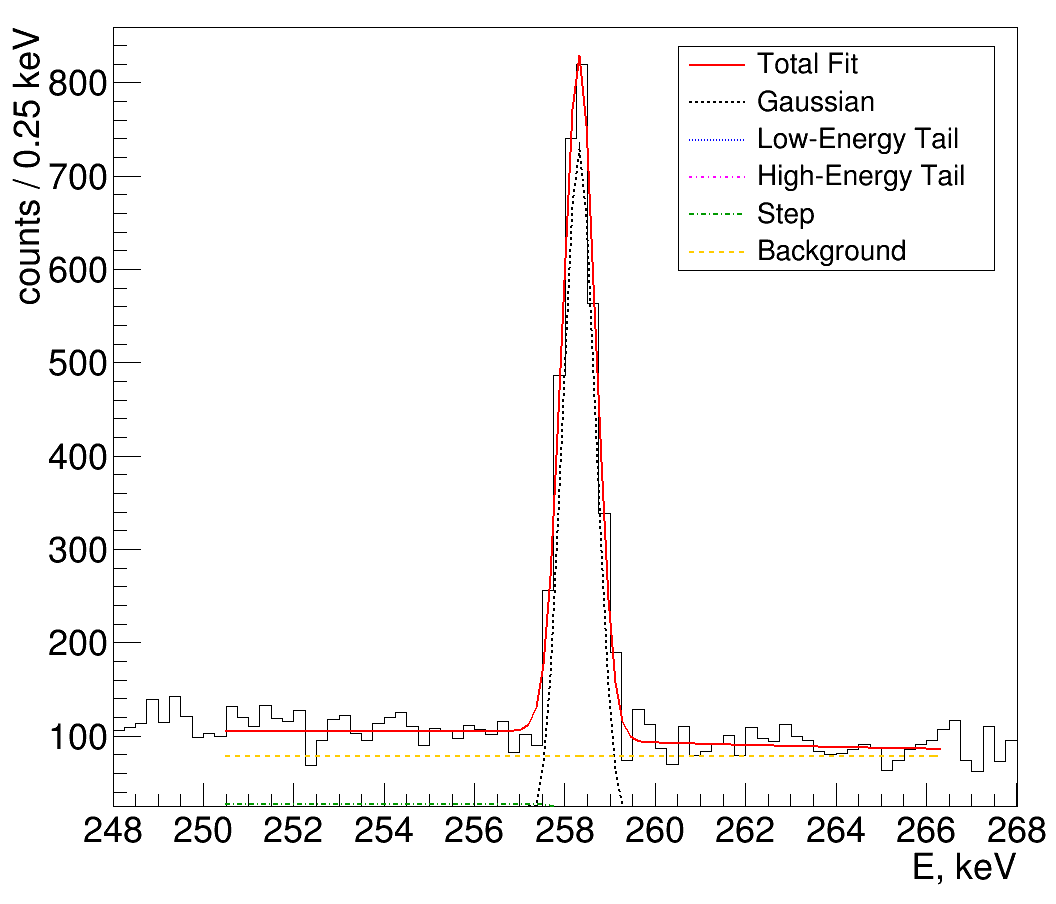}
    \caption{} 
    \label{fig:fit_a}
  \end{subfigure}
  \hfill
  \begin{subfigure}{0.45\textwidth}
  \centering
    \includegraphics[width=\linewidth]{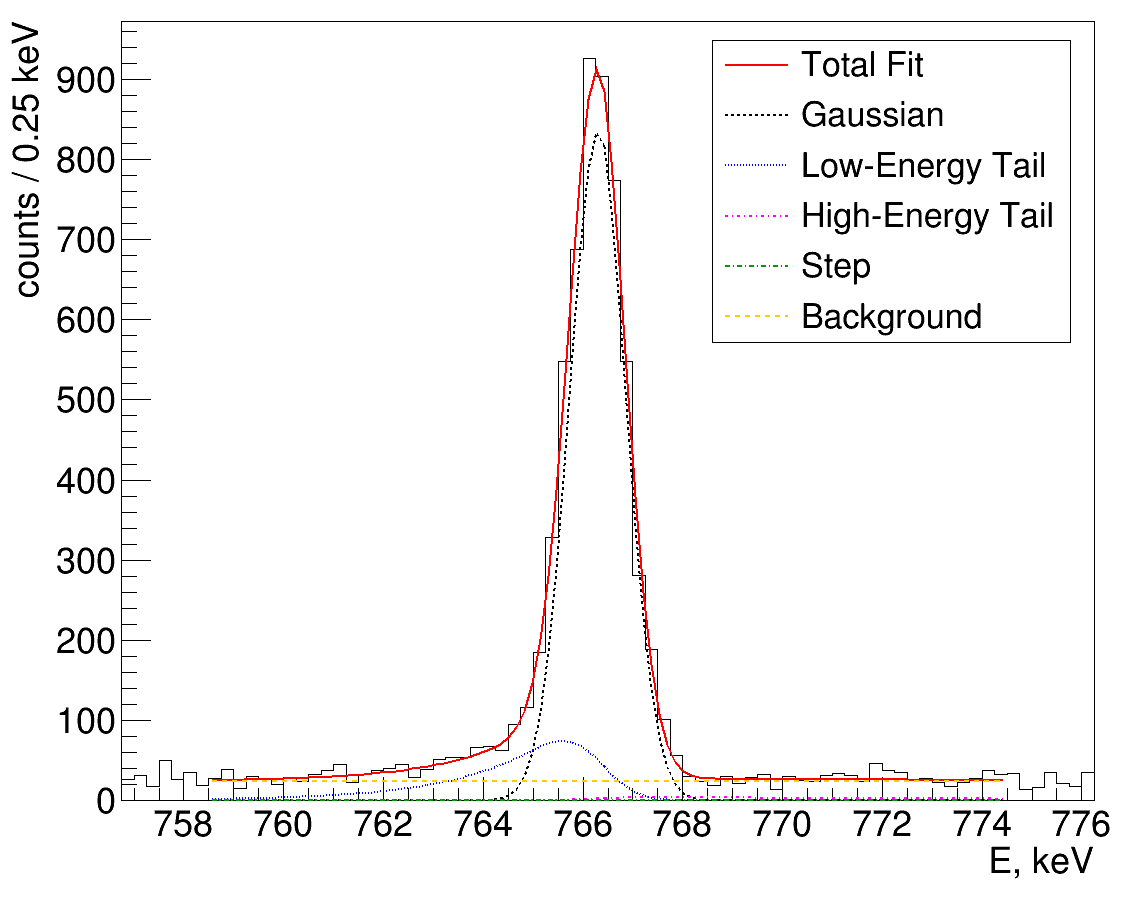}
    \caption{} 
    \label{fig:fit_b}
  \end{subfigure}
  \caption{Fit of the peaks with function from \cite{arnquist2023energy}.}
  \label{fig:fit}
\end{figure}

For the effective mass estimation we compared 766 keV $\gamma$-line from experiment and Monte-Carlo, while 258 keV line was used for a cross-check. Moreover, 766 keV $\gamma$-quanta lose their energy more homogeneously in the detector volume, than the 258 keV $\gamma$-quanta, which lose more energy closer to the top part of the crystal. In order to assess systematic effects, two approaches were applied to determine the counts in the 258 keV and 766 keV peaks. First, the peaks were fitted with the function (\ref{eq:fit}) described in \cite{arnquist2023energy} that was used in the MAJORANA DEMONSTRATOR data analysis 

\begin{equation}
PS(E)=G(E)+T_{LE}+T_{HE}+S(E)
\label{eq:fit}
\end{equation}

\noindent where $G(E)$ is  Gaussian function; $T_{LE}$ and $T_{HE}$ are exponentially modified Gaussian tail functions; $S(E)$ is step background. The fitting results are shown in Figure \ref{fig:fit}.

\begin{figure}[H] 
  \centering
  \begin{subfigure}{0.45\textwidth}
  \centering
    \includegraphics[width=0.88\linewidth]{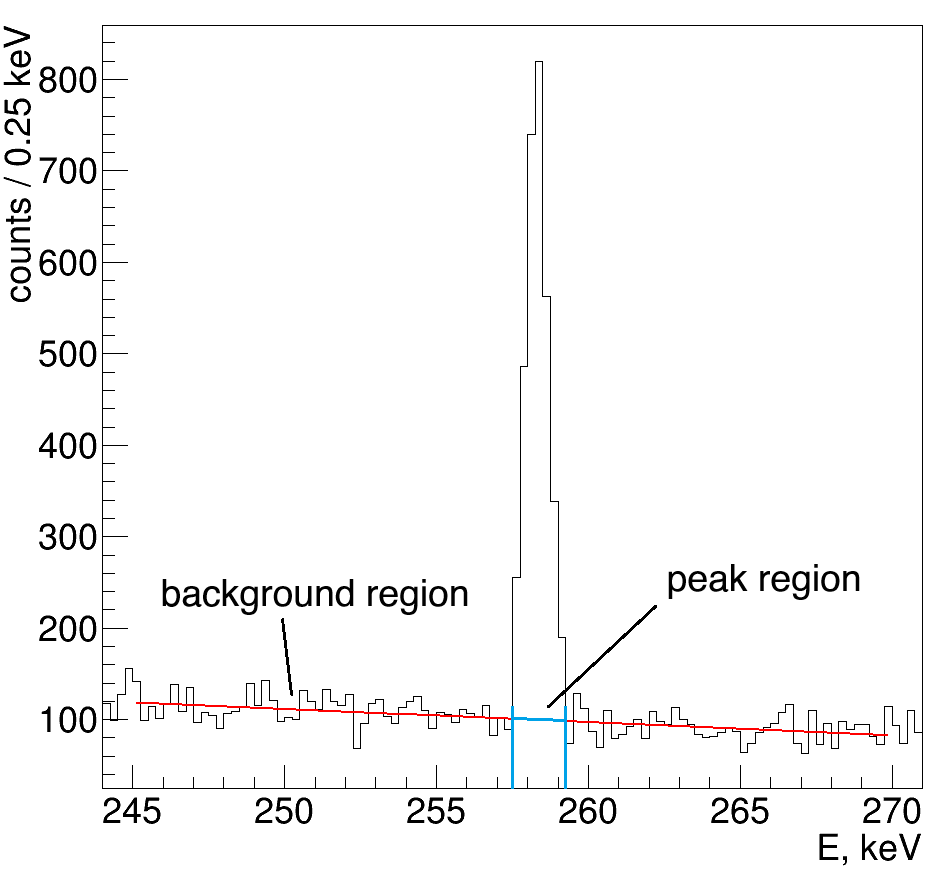}
    \caption{} 
    \label{fig:simple_a}
  \end{subfigure}
  \hfill
  \begin{subfigure}{0.45\textwidth}
  \centering
    \includegraphics[width=\linewidth]{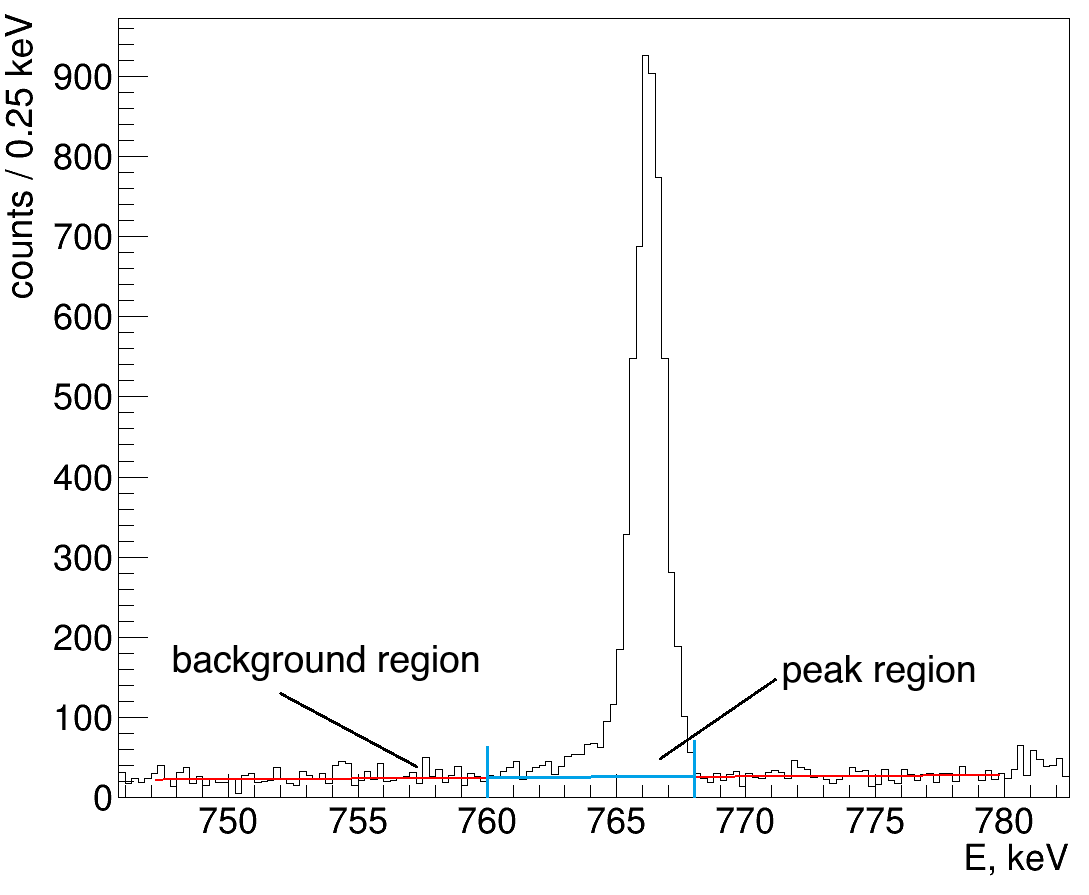}
    \caption{} 
    \label{fig:simple_b}
  \end{subfigure}
  \caption{Fit of the peaks with “simple” method.}
  \label{fig:simple}
\end{figure}

As a cross-check, the counts were evaluated using a “simple” method: the continuum under the peak was approximated by the straight line, and the net counts under the peak were obtained after subtracting this background. The results of this method are shown in Figure \ref{fig:simple}.

The results obtained from both methods were summed for each peak. They are presented in Tables \ref{tab:fit} and \ref{tab:simple} and compared with results of both Monte Carlo simulations (MC$_1$ is 0.06 mm dead layer and MC$_2$ is 0.5 mm dead layer. The uncertainty in the experimental peak shape results in larger discrepancies for the 258 keV line with a higher background level.

\begin{table}[H]
\centering
\caption{Comparison of counts (fitting method).}
\label{tab:fit}
\begin{tabular}{cccccc}
\hline
$E$ (keV) & Experiment & MC$_1$ & MC$_2$& Experiment/MC$_1$ & Experiment/MC$_2$ \\
\hline
258.26 & $2797 \pm 62$ & 2578 & 2583 & $1.08 \pm 0.02$ & $1.08 \pm 0.02$\\
766.36 & $5688 \pm 108$ & 5941 & 5946 & $0.96 \pm 0.02$ & $0.96 \pm 0.02$\\
\hline
\end{tabular}
\end{table}

\begin{table}[H]
\centering
\caption{Comparison of counts (simple method).}
\label{tab:simple}
\begin{tabular}{cccccc}
\hline
$E$ (keV) & Experiment & MC$_1$ & MC$_2$ & Experiment/MC$_1$ & Experiment/MC$_2$ \\
\hline
258.26 & $2637 \pm 76$ & 2578 & 2583 & $1.02 \pm 0.02$ & $1.02 \pm 0.02$ \\
766.36 & $5735 \pm 88$ & 5941 & 5946 & $0.97 \pm 0.02$ & $0.96 \pm 0.02$\\
\hline
\end{tabular}
\end{table}

 Estimated effective mass of HPGe crystal is $1.37 \pm 0.07$ kg, which is in agreement with a value from manufacturer’s specification.. The main systematic uncertainty originates from the limited knowledge of the exact crystal position inside the detector housing, including the distance to the endcap and incorrect internal endcap dimensions. In contrast, effects related to source positioning and dead-layer thickness are small (<1–2\%) and do not significantly affect the mass result. One can consider $\sim$5\% as an additional uncertainty in the $\nu$GeN experiment related to the effective mass of the crystal. This is highly important because the value is significantly lower when compared to the main ambiguities which are due to the unknown quenching, the background model, and the intensity of the high-energy part of the antineutrino spectrum. Increasing the precision of future neutrino experiments with HPGe detectors will demand a better understanding of the detector's effective mass. This requires precise knowledge of the dead layer properties: its thickness, homogeneity, and evolution over time. If needed, such parameters can be constrained through dedicated calibration measurements.
 
\section{Conclusion}
As a result of the calibration measurements performed with the distributed 
$^{238}$U source of known activity (accuracy at the level of 0.5\%), the conformity of the mass of the HPGe detector used in the $\nu$GeN experiment with its nominal specification has been verified.

The applicability of the calibration method using a distributed source for determining the mass and volume of germanium detectors has been demonstrated.

It is worth mentioning that the above calibrations will also be applied to measure the samples enriched in $^{96}$Zr which were produced for an experiment searching for rare decay modes of this nucleus.

\section{Acknowledgments}
This work is supported by the State Project "Science" by the Ministry of Science and Higher Education of the Russian Federation under Contract 075-15-2024-541.




\end{document}